\documentclass[fleqn,usenatbib,nofootinbib]{mnras}

\usepackage{newtxtext,newtxmath}

\usepackage[T1]{fontenc}

\DeclareRobustCommand{\VAN}[3]{#2}
\let\VANthebibliography\thebibliography
\def\thebibliography{\DeclareRobustCommand{\VAN}[3]{##3}\VANthebibliography}

\usepackage{graphicx}	
\usepackage{amsmath}	
\usepackage{longtable}
\usepackage{booktabs} 
\usepackage{multirow}
\usepackage{todonotes} 
\usepackage{adjustbox}
\usepackage{booktabs} 
\usepackage{longtable}
\usepackage{lscape}
\usepackage{array}
\makeatletter
\renewcommand{\@journal}{}
\makeatother

\title[Fast Radio Bursts as cosmological proxies]{Fast Radio Bursts as cosmological proxies: estimating the Hubble constant}

\author[E. F. Piratova-Moreno et al.]{\large 
Eduard Fernando Piratova-Moreno,$^{1}$\thanks{E-mail: efpiratovam@libertadores.edu.co}
Luz Ángela García,$^{2}$
Carlos A. Benavides-Gallego,$^{3}$
Carolina Cabrera$^{1}$
\\
$^{1}$Fundaci\'on Universitaria Los Libertadores. Cra. 16 No. 63A-68, Bogot\'a, Colombia, C\'odigo Postal 111221\\
$^{2}$Universidad ECCI, Cra. 19 No. 49-20, Bogot\'a, Colombia, C\'odigo Postal 111311\\
$^{3}$School of Physics and Astronomy, Shanghai Jiao Tong University, 800 Dongchuan Road,
Minhang, Shanghai 200240, PRC.
}

\pubyear{2015}

\begin{document}
\label{firstpage}
\pagerange{\pageref{firstpage}--\pageref{lastpage}}
\pagestyle{plain}
\maketitle

\begin{abstract}
One of the most challenging problems in modern cosmology is the Hubble tension, a discrepancy in the predicted expansion rate of the Universe with different observational techniques that results in two conflicting values of $H_0$. We leverage the sensitivity of the Dispersion Measure (DM) from Fast Radio Bursts (FRBs) with the Hubble factor to investigate the Hubble tension. We build a catalog of 98 localized FRBs and an independent mock catalog and employ three methods to calculate the best value of the Hubble constant: i) the mean of $H_0$ values obtained through direct calculation, ii) the maximum likelihood estimate (MLE), and iii) the reconstruction of the cosmic expansion history $H(z)$ using two DM-$z$ relations previously explored in the literature.  When the dataset of confirmed FRBs is employed, our predictions are compatible with reports from the Planck collaboration 2018, with $H_{0} =  65.13 \pm 2.52 \, \text{km/s/Mpc}$ and $57.67 \pm 11.99 \, \text{km/s/Mpc}$ for maximum likelihood and the arithmetic mean, respectively. On the other hand, if we assume a linear and a power-law function for the DM-$z$ relation, our predictions for $H_{0}$ are $51.27 ^{+3.80}_{-3.31} \, \text{km/s/Mpc}$ and $77.09^{+8.89}_{-7.64} \, \text{km/s/Mpc}$, respectively. Finally, using 100 mock catalogs of 500 simulated FRBs in each realization, we obtain larger values for $H_0$ with all methods considered: $H_{0;\text{ Like}} = 67.30 \pm 0.91 \, \text{km/s/Mpc}$, $H_{0;\text{ Mean}} = 66.21 \pm 3.46  \, \text{km/s/Mpc}$, $H_{0;\text{ Median}} = 66.10 \pm 1.89  \, \text{km/s/Mpc}$, $H_{0;\text{ Linear}} = 54.34 \pm 1.57 \, \text{km/s/Mpc}$ and $H_{0;\text{ Power-law}} = 91.84 \pm 1.82 \, \text{km/s/Mpc}$ for the MLE method, the arithmetic mean, and linear and power-law $\text{DM}-z$ relations, respectively. More importantly, our results for mock FRB catalogs notably increase the statistical precision, ranging from 1.4\% to 5.2\% for the MLE method and arithmetic mean. In particular, our result with the MLE applied to synthetic FRBs is at the same level of precision as reports from SH0ES. Ultimately, the rapid increase in the number of confirmed FRBs will provide us with a robust prediction of the value of the Hubble constant, which, in combination with other cosmological observations, will allow us to alleviate (to a certain degree) the current Hubble tension.
\end{abstract}




\section{Introduction}
Fast Radio Bursts (FRBs) are bright millisecond transients detected in the radio waveband, ranging from 100 MHz to 8 GHz. Although their origin has not yet been discovered, nor the mechanisms that fuel these short bursts of energy, they have been depicted as having cosmological origin; therefore, they could be used to identify the properties of their galaxy hosts and the intergalactic medium through which their light goes before reaching our telescopes.\newline
Besides the 2D angular coordinates, the main observable used to study FRBs is their dispersion measure (DM), which quantifies the properties of the medium where their light passes through, more specifically, the integrated number of free electrons along the path that interacts with their photons and causes a difference in the frequencies of light emitted by the object. This interaction between the FRB's photons and free electrons causes a difference between the radiation's highest and lowest frequencies, allowing us to infer the total DM. \newline
Different authors have claimed that three components contribute to the total DM$_{\text{obs}}$ \cite{pol2019,zhang2020,zhu2021}, among them, the dispersion due to our galaxy, DM$_{\text{MW}}$, and due to their extragalactic origin, a term due to the intergalactic medium (IGM), DM$_{\text{IGM}}$, and a third one caused by the interaction of the FRB photons with their host galaxy, DM$_{\text{host}}$. The second and third terms have intricate dependencies on redshift $z$, explained mainly by their cosmological origins. Although the modeling of each one of the latter terms is complex and constitutes an extensive field of study of FRBs.\newline

To date, a few thousand FRBs have been detected, and the census is rapidly growing with the new radio telescopes that come online that observe and characterize these transients. Only a tiny fraction of these reported FRBs in the literature has direct information about their $z$ because their galactic progenitors cannot be identified with radio observations; thus, other wavelength surveys need to observe the patch of the sky where the FRB is detected to have an inferred $z$ and complete their localization. However, less than a hundred FBRs have been detected with known $z$ (i.e., their hosting galaxy is identified), many of which are at low redshift. \newline
Recent studies focus on modeling the relationship between the observed DM and their corresponding $z$. For instance, \cite{macquart2020,cui2022,baptista2023} presented different fits for a linear relation in DM vs. $z$. In particular, \cite{piratova2024} presents a revision of different DM$_{\text{obs}}$ models as a function of the redshift based on physical motivations. In the latter work, the authors used 24 FRBs with $z$ known to propose different relations for DM$_{\text{obs}}$($z$): a linear trend, a log parabolic relation, a power-law model, and as a separate stage, an interpolation that not only takes into account $z$, but also the angular coordinates of each FRB, following results from \cite{xu2021}.\newline

On the other hand, FRBs could offer a new window for cosmological studies because of their random distribution in the sky. Eventually, it will allow us to cross-correlate their position with galaxy redshift surveys. Despite the increasing number of cosmological FRBs, those with $z \geq$ 0.5, there is an inner challenge in deriving cosmological parameters and delivering information about the state of the intergalactic medium through the study of the dispersion measure of these transients due to the small sample of identified FRBs. Nonetheless, astronomers can use a combined analysis with other cosmological observations, such as the luminosity distance of SNIa, inferred cosmology from the cosmic microwave background (CMB), Gamma Ray Bursts (GRB), Big Bang Nucleosynthesis (BBN), Baryon Acoustic Oscillations (BAO), among others, to derive solid constraints on the value of the fine-structure constant, the value of the baryon fraction $\Omega_b$ (given the correlation with the number of free electrons calculated from the DM), the equation of state of the dark energy 
\cite{deng2014,zheng2014}, and $H_0$ (the Hubble constant measured today). Notably, FRB studies provide an independent method to calculate a value of $H_0$, which would benefit the astronomical community currently facing the so-called ``Hubble tension'' \cite{divalentino2021,review2023,verde2024}. The Hubble tension invokes a large discrepancy between the values of $H_0$ inferred from the CMB: 67.4 $\pm$ 0.5 km/s/Mpc \cite{planck2018}, and a recent release of SNIa: 73.0 $\pm$ 1.0 km/s/Mpc \cite{riess2022}, measured in the local universe from the SH0ES collaboration. Since the dispersion measure of the FRBs accounts for the distance between the transient and us, the DM is related to $H_0$ through the Hubble parameter in a non-trivial way and, thus, can be used as a proxy to estimate $H_0$. \newline

Different groups have started a campaign to derive values of $H_0$ based on the current FRBs available. For instance, \cite{hagstotz2022} reported a value of 62.3 $\pm$ 9.1 km/s/Mpc using 12 FRBs. On the other hand, \cite{wu2022} presented a final value of $H_0 =$ 68.81$^{+4.99}_{-4.33}$ km/s/Mpc using 18 localized FRBs with a theoretical approach. With the same number of FRBs, \cite{Liu_2023} reported a value of 71$\pm 3\,$km/s/Mpc. Using 16 FRBs observed with ASKAP, \cite{james2022} reported a value of 73$^{+13}_{-8}$ km/s/Mpc. Moreover, \cite{zhang2022} reported a $H_0$ of 80.4$^{+24.1}_{-19,4}$ km/s/Mpc for the expansion rate today with 12 unlocalized FRBs and BBN constraints. Furthermore, \cite{wei2023} provided an estimate  $95.8^{+7.8}_{-9.2}\,$ km/s/Mpc, accounting for 24 reported FRBs with known $z$. \cite{hoffmann2024} reported a value for $H_0$ of 64$^{+15}_{-13}\,$km/s/Mpc, making use of 26 selected FRBs from the DSA, FAST and CRAFT surveys. On the other hand, \cite{refId0} reported $H_0= 74^{+7.5}_{-7.2}\,$km/s/Mpc using 30 FRBs and the temporal scattering of the FRB pulses due to the propagation effect through the host galaxy plasma. Applying Monte Carlo simulation to 69 localized FRBs,  \cite{Gao_2024kkx} presented a value of Hubble constant $70.41^{+2.28}_{-2.34}\,$km/s/Mpc. Finally, \cite{wang2025} performed an analysis combining 92 localized FRBs with data for DESI Y1, indicating a preference for a dynamical dark energy model $\omega_0$-$\omega_a$ over the $\Lambda$CDM standard model, and with the FRB subset-only, and values for $H_0$ of 69.04$^{+2.30}_{-2.07}$ km/s/Mpc and 75.61$^{+2.23}_{-2.07}$ km/s/Mpc, assuming galactic electron density models NE2001 \cite{cordes2002} and YMW16 (Yao et al. in prep), respectively.\newline 

In a similar fashion, we extend the catalog of FRBs presented in \cite{piratova2024} to the latest 98 localized FRBs in the literature to perform a thorough statistical analysis and find a robust value for $H_0$. We use the maximum likelihood estimate (MLE) method and other statistical metrics to find the best value of $H_0$ with the FRBs in the observed catalog. Then we calculate the value of $H_0$ assuming two of the models explored in \cite{piratova2024}: the linear trend and the power-law function of the DM$_{\text{obs}}$ with $z$ and follow the evolution of H$(z)$, as explored in \cite{fortunato2024}. Finally, we build a synthetic catalog of FRBs (increasing the observed sample by a factor of 5) to derive the best-fit value of $H_0$ with each one of the methods described with confirmed FRBs.\newline

The outline of this work goes as follows: section~\ref{DM} describes in detail how the different contributions to the dispersion measure (DM) are modeled. Section~\ref{sec:analysis} is devoted to presenting our statistical analysis to retrieve the best values of $H_0$ implementing three completely different methods and, the confirmed FRBs. Section~\ref{mocks} shows the analysis made with mock catalogs and presents the values obtained for the Hubble constant in this way. Finally, section~\ref{sec:concl} presents the main conclusions found in this work. Throughout the paper, we have assumed a flat $\Lambda$CDM model with cosmological parameters from \cite{planck2018}.

\section{Dispersion Measure (DM)}\label{DM}

As Fast Radio Burst (FRB) signals travel toward Earth, they undergo dispersion, resulting in a time delay between the arrival of different frequencies. This delay is quantified using the dispersion measure (DM) as follows:\newline
\begin{align}
\Delta t \propto \left( \nu_{\text{lo}}^{-2} - \nu_{\text{hi}}^{-2} \right)\cdot\text{DM},
\end{align}
with $\nu_{\text{lo}}$ and $\nu_{\text{hi}}$ are the lowest and highest frequencies of the emitted signal, respectively. The dispersion measure is associated with the column density of free electrons along the signal's path and is expressed as:
\begin{align}
\text{DM} = \int \frac{n_e}{(1 + z)} \, dl,
\end{align}
where $n_e$ is the cosmic free electron density, and $l$ is the line of sight to the FRB. The observed dispersion measure ($\text{DM}_{\text{obs}}$) is estimated from the following contributions: 
\begin{align}\label{obs}
\text{DM}_{\text{obs}} =   \text{DM}_{\text{IGM}} + \text{DM}_{\text{MW}} +\frac{\text{DM}_{\text{host}}}{1+z},
\end{align}
where the dispersion measure due to the Milky Way  is $ \text{DM}_{\text{MW}}$;  $\text{DM}_{\text{IGM}}$, the contribution of the intergalactic medium to the dispersion; and $\text{DM}_{\text{host}}$, corresponds to the contribution 
of the host galaxy.

\subsection{Dispersion measure due to the Galaxy Host ($\text{DM}_{\text{host}}$) and Milky Way ($\text{DM}_{\text{MW}}$)}

We assume that the contribution from the host galaxy follows a similar form to that of the Milky Way, as:
\begin{align}
\text{DM}_{\text{host}} = \frac{100}{1 + z_{\text{host}}}\text{\, pc\,cm$^{-3}$}\, , 
\end{align}
with an associated uncertainty given by \cite{hagstotz2022}:
\begin{align}
\sigma_{\text{host}} =   \frac{50}{1 + z_{\text{host}}} \text{\, pc\,cm$^{-3}$}.
\end{align}
On the other hand, the contribution of the Milky Way ($\text{DM}_{\text{MW}}$) has been estimated to range between 50-100 $\mathrm{pc\,cm^{-3}}$ \cite{stz261}, with a value of 100 $\mathrm{pc\,cm^{-3}}$ selected for this work. The predicted DM values align with pulsar observations within an uncertainty of approximately 
$ \sigma_{\text{MW}} \sim 30 \, \mathrm{pc \, cm^{-3}}$ \cite{Manchester_2005}, which is adopted as the model's uncertainty estimate.

\subsection{Intergalactic Medium Dispersion Measure ($\text{DM}_{\text{IGM}}$)}

The contribution of the intergalactic medium to the dispersion measure can be expressed as follows:
\begin{align}
\text{DM}_{\text{IGM}}(z) = \int_0^z n_e(z') f_{\text{IGM}}(z') \frac{1 + z'}{H(z')} \, dz',
\end{align}
\noindent where $H(z)$ is the Hubble parameter that intrinsically depends on $H_0$ and $z$, and $f_{\text{IGM}}$, the baryon mass fraction. For the redshift range considered, where almost all baryons are ionized, the cosmic electron density can be approximated as a function of the baryon abundances: $n_e(z) \approx \chi_e\frac{\bar{\rho_b}}{m_p}$ \cite{hagstotz2022}, where $\bar{\rho_b}$ is the mean baryon density, $m_p$ the proton mass, and the electron fraction $\chi_e = Y_H +\frac{1}{2}Y_{He}$. Assuming that the primordial abundances of Hydrogen and Helium satisfy the relationships $Y_H\approx 1-Y_{He}$ and $Y_{He}=0.24$, as determined by CMB measurements \cite{2020A&A...641A...6P} and spectroscopic observations \cite{Aver_2015}, $\chi_e \approx 1-\frac{1}{2}Y_{He}$. On the other hand, the fraction of electrons in the intergalactic medium, $f_{\text{IGM}}$, is calculated by subtracting the fraction bound in stars, compact objects, and dense interstellar medium (ISM). For the present analysis, we adopt $f_{\text{IGM}} = 0.84$, following \cite{hagstotz2022}.

With the previous considerations, the IGM contribution can be written as follows:
\begin{align}\label{IGM}
\text{DM}_{\text{IGM}}(z) =  \frac{3c \Omega_b H_0}{8 \pi G m_p} \, \chi_e \, f_{\text{IGM}} \int_0^z  \frac{1 + z'}{E(z')} \, dz'\, .
\end{align}
Here, $\Omega_b$ is the dimensionless baryon density parameter and 
$E(z)$ is the reduced Hubble factor defined as $E(z)=\frac{H(z)}{H_0}$, under the standard $\Lambda$CDM model. Since eq.~\eqref{IGM} has a direct dependence on $H_0$, we can use it to calculate the value of this cosmological parameter.\newline
Finally, hydrodynamic simulations suggest an inhomogeneous distribution of electrons in the IGM, such a distribution can be effectively modeled by a Gaussian centered around the value provided by the equation~\eqref{IGM}. On the other hand, a linear interpolation of the uncertainties is applied between the values reported by the simulations, specifically 
$ \sigma_{\text{IGM}}(z=0) \approx 40\,\text{pc\,cm}^{-3}$ and  $\sigma_{\text{IGM}}(z=1) \approx 180\,\text{pc\,cm}^{-3}$ \cite{hagstotz2022}.\\

\section{Calculating $H_0$ with our FRB  catalog}\label{sec:analysis}

This section presents three methods for estimating 
$H_0$ using localizated FRB data in Table~\ref{data}. FRB221027A is excluded for its ambiguity in host galaxy localization, so we calculate with 97 FRBs. In the first method, the mean value obtained for $H_0$ is based on the sensitivity of the IGM term to this parameter. 
The second method focuses on maximizing the likelihood function, while the third one derives from an expression for the cosmic expansion history, $H(z)$, and evaluates it at $z = 0$.

\subsection{$H_0$ derived with the arithmetic mean}

\begin{figure}
\includegraphics[width=9cm]{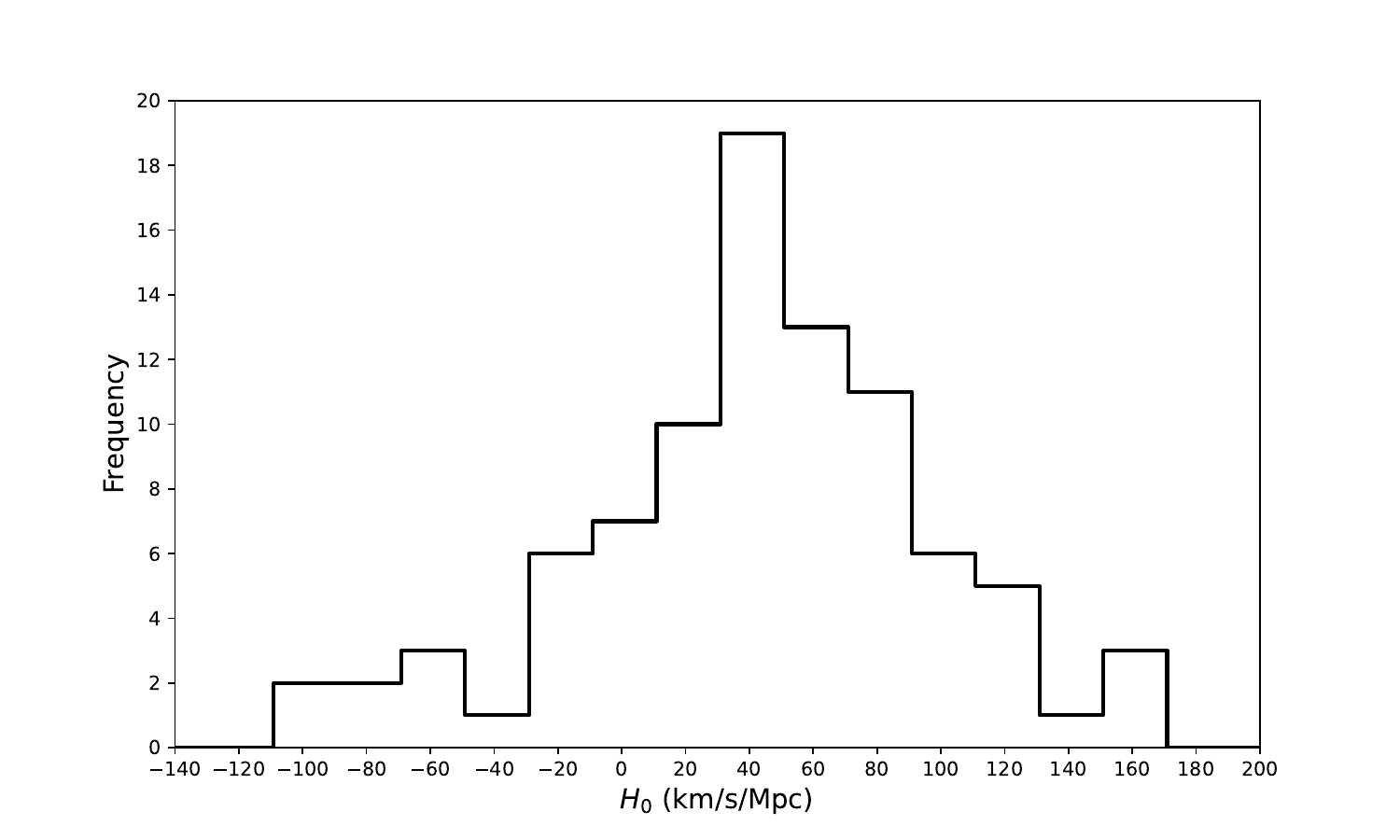}
\caption{\tiny{The histogram represents the frequency distribution of $H_0$ values derived from the data. Using the arithmetic mean, the value for the Hubble constant with 97 confirmed FRBs is $H_0 = 57.67 \pm 11.99 \,\text{km/s/Mpc}.$
}}
\label{fig_histogram}
\end{figure}

\begin{figure}
\begin{center}
\includegraphics[width=9cm]{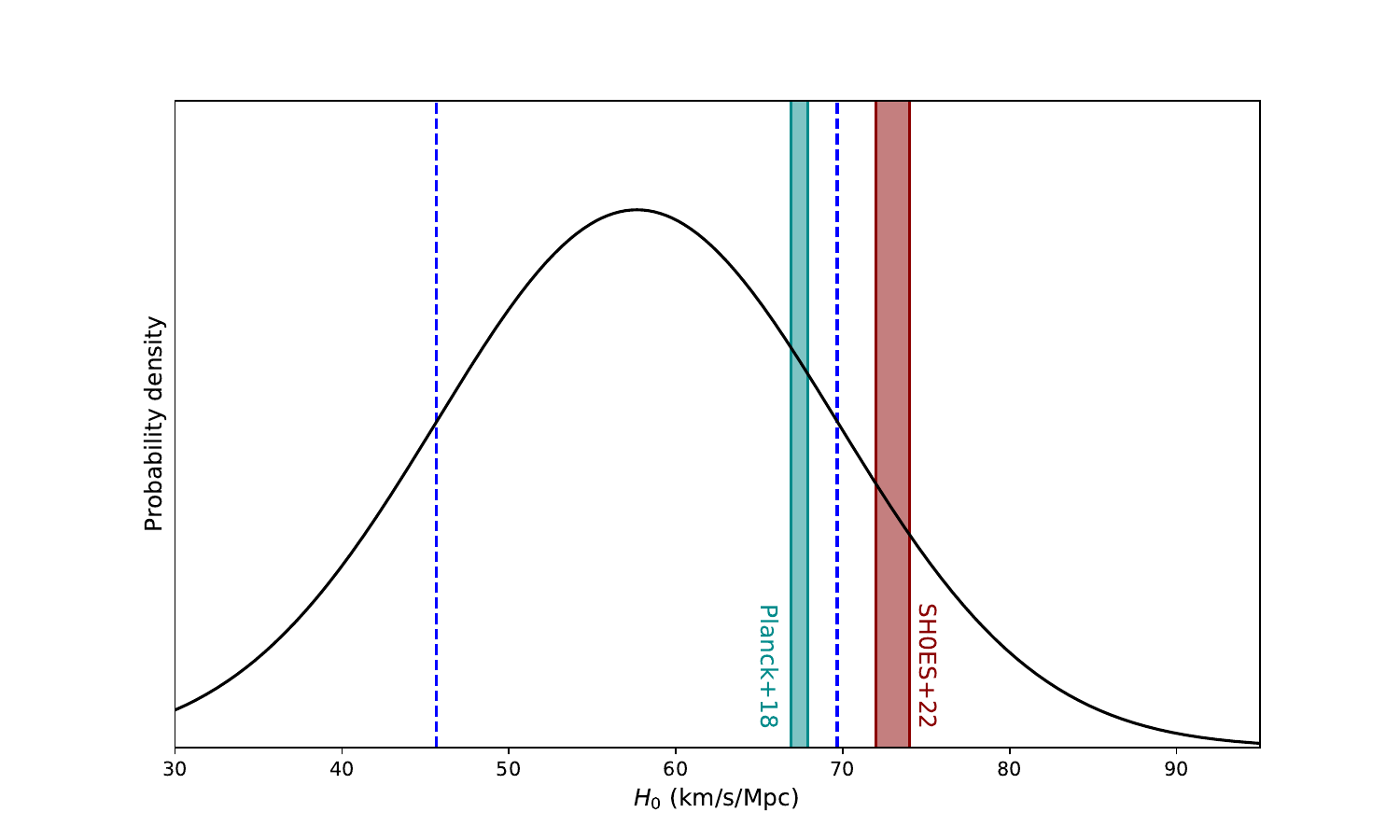}
\end{center}
\caption{The Gaussian distribution obtained from the direct calculation and its associated errors is compared with the values reported by the \citet{planck2018} and \citet{riess2022} measurements. The blue dotted lines are the errors estimated as one standard deviation.}\label{gauss_direct}
\end{figure}

We can use equation (7) to calculate $H_0$ for each localized FRB presented in Table \ref{data} and then obtain the arithmetic mean of the data set. To this purpose, we substitute each contribution to DM discussed in Section \ref{DM} into eq.~\eqref{obs}, such that:
\begin{align}\label{DMobsMean}
\text{DM}_{\text{obs}} = \frac{3c \chi_e f_{\text{IGM}} \cdot 10^4 \Omega_b h^2}{8 \pi G m_p H_0} \int_0^z \frac{1 + z'}{E(z')} \, dz' + \text{DM}_{\text{MW}} + \frac{\text{DM}_{\text{host}}}{1 + z}.
\end{align}
From eq.~\eqref{DMobsMean}, $H_0$ can be derived as:
\begin{align}
H_0 = \frac{3c \chi_e f_{\text{IGM}}}{8 \pi G m_p}  \frac{10^4 \Omega_b h^2}{\text{DM}_{\text{obs}} - \text{DM}_{\text{MW}} - \frac{\text{DM}_{\text{host}}}{1+z}} \int_0^z \frac{1 + z'}{E(z')} \, dz'.
\end{align}
The computed values of $H_0$ with 97 FRBs in Table~\ref{data} are displayed in Figure~\ref{fig_histogram}.
For each presented value in the histogram, we compute the error using error propagation as follows:
\begin{equation}\label{errorMean}
 \sigma_{H_0} = \sqrt{\sum_{i=1} \left( \frac{\partial H_0}{\partial x_i} \cdot \sigma_{x_i} \right)^2},   
\end{equation}
where $x_i$ represents each of the variables that contribute to the propagation of errors and $\sigma_{x_i}$ its errors, which in this case are: $\text{DM}_{\text{obs}}$, $\text{DM}_{\text{MW}}$, $\text{DM}_{\text{host}}$ and $z$. The value for $H_0$ obtained through this method is:
\begin{equation}\label{H0mean}
H_0 = 57.67 \pm 11.99 \, \text{km/s/Mpc}.
\end{equation}
Figure~\ref{gauss_direct} displays the result in equation \eqref{H0mean} in addition with the $H_0$ values reported from the \cite{planck2018} and SH0ES measurements \cite{riess2022}. Up to a precision of $\sigma = 11.99\,\text{km/s/Mpc}$, our result is compatible with \cite{planck2018} data, and excludes in 1$\sigma$ SH0ES measurements \cite{riess2022}.  However, as evident from Figure~\ref{gauss_direct}, a statistical precision of $\approx$ 21\% is extremely low. Nevertheless, as we will demonstrate later with mock catalogs, it is expected that with a few hundred data points, our predictions for $H_0$ and its statistical precision will improve, and our forecast of this cosmological parameter will be more consistent with the values reported by the referred collaborations.

\subsection{$H_0$ from cosmic expansion history}
In this section, we discuss and apply a method to reconstruct the cosmic expansion history through the Hubble parameter $H(z)$ and estimate its value today, i.e., $H_0$. We remind the reader that we use the sample of localized FRBs in Table~\ref{data}, and, models first introduced in \cite{piratova2024}.

The Hubble parameter, redshift, and the mean dispersion measure of IGM for FRBs are interrelated through the equation \eqref{IGM}. We obtain a explicit expression for $H(z)$ by deriving it \cite{fortunato2024}:
\begin{align}\label{Hz}
H(z) = \frac{3c 10^4 \Omega_b h^2}{8 \pi G m_p}  \chi_e f_{\text{IGM}}  (1 + z)  \left( \frac{d  \text{DM}_{\text{IGM}}}{dz} \right)^{-1}.
\end{align}
\noindent Equation~\eqref{Hz} can be used to reconstruct the cosmic expansion history, $H(z)$, considering the derivative of $\text{DM}_{\text{IGM}}$ with respect to $z$ and using the latter to estimate $H_0$ as shown in \cite{Liu_2023} and more recently in \cite{fortunato2024}. The models presented in \cite{piratova2024} for 24 known FRBs (at the moment) have been updated to 97 in Table~\ref{data}. Among the models originally discussed in \cite{piratova2024}, only two have been considered in this work due to their superior performance when predicting $z$ from the observed DM: the linear and power-law models. From eq.~\eqref{obs},  $\text{DM}_{\text{IGM}}$ is given by:
\begin{align}
\text{DM}_{\text{IGM}}  &=  \text{DM}_{\text{obs}}  - \text{DM}_{\text{MW}} - \frac{\text{DM}_{\text{host}}}{1+z}\, 
\end{align}

If the relationship between $\text{DM}_{\text{obs}}$ and $z$ is linear, as the Macquart relation  \cite{macquart2020, piratova2024}:
\begin{align}\label{linear_dm}
\text{DM}_{\text{obs}} &= az + b, 
\end{align}
\noindent with our sample of 97 FRB data points, the best-fit parameters for the linear trend are: $a = 959.32\pm 73.12 \text{\, pc\,cm$^{-3}$}$  and $b = 240.11 \pm 25.67 \text{\, pc\,cm$^{-3}$}$.\newline

Now, if the DM-$z$ relationship is a the power-law function \cite{piratova2024}, $\text{DM}_{\text{obs}}$ is modeled as:
\begin{align}\label{pow_dm}
\text{DM}_{\text{obs}} &= A(1+z)^{\alpha}.
\end{align}
\noindent The updated values of $A$ and $\alpha$ our 97 localized FRBs are: $A = 297.25 \pm 17.90\text{\, pc\,cm$^{-3}$}$, and $\alpha = 2.03 \pm 0.13$. \newline

Using the relationships \eqref{linear_dm} and \eqref{pow_dm} and the updated values for the free parameters in each case, the derivative $\frac{dDM_{\text{IGM}}}{dz}$ can be obtained and substituted into the eq.~\eqref{Hz}. The Hubble parameter with our linear model is given by:
\begin{align}\label{HzLineal}
\frac{H(z)}{1 + z} = \frac{3c}{8 \pi G m_p}  10^4 \Omega_b h^2    f_{\text{IGM}}(z)  f_e(z)  \left[ 959.32 + \frac{100}{(1 + z)^2}\ \right]^{-1},
\end{align}
\noindent and equivalently, for our power-law function:
\begin{align}\label{HzPoweLaw}
\frac{H(z)}{1 + z} = \frac{3c}{8 \pi G m_p}  10^4 \Omega_b h^2   f_{\text{IGM}}(z)  f_e(z)  \left[ 604.48 (1 + z)^{1.03} + \frac{100}{(1 + z)^2} \right]^{-1}
\end{align}

By setting $z=0$ in equations \eqref{HzLineal} and \eqref{HzPoweLaw}, we recover a value for the Hubble constant today $H_0 =51.27 ^{+3.80}_{-3.31}\,\text{km/s/Mpc}$ for the linear model, and $H_0 = 77.09^{+8.89}_{-7.64}\,\text{km/s/Mpc}$ for the power-law function, with statistical precisions of 7.4\% and 11.5\%, respectively.\newline
In this case, the prediction of the linear model is inconsistent both with \cite{planck2018} and  \cite{riess2022}, but privileges lower values of $H_0$. On the other hand, the prediction of the power-law model is comparable with SH0ES measurements within the error range. Interestingly, the linear trend (also known as the Macquart relation) is extensively applied at the low $z$ regime, but it leads to a poor prediction of $H_0$ with our current census of FRBs.

\subsection{$H_0$ from Maximum Likelihood Estimate (MLE).}

 This section is devoted to estimating the best-fit value of $H_0$ when we maximize the Likelihood function with data in Table~\ref{data}. The likelihood function quantifies how well a given model of $H_0$ explains the observed data, accounting for uncertainties in the measurements. This function is defined as:
\begin{equation}
    \ln \mathcal{L}(H_0) = -\frac{1}{2} \sum_{i} \left[ \frac{(\text{DM}_{\text{obs};i} - \text{DM}_{\text{model};i})^2}{\sigma_i^2} + \ln(\sigma_i^2) \right]
\end{equation}
where $\text{DM}_{\text{obs}}$ represents the total dispersion measure, $\text{DM}_{\text{model}}$ is the theoretical prediction dependent on $H_0$, and $\sigma$ encompasses the total uncertainty, including contributions from the host galaxy, intergalactic medium, and Milky Way shown in section~\ref{DM}.\newline
Our computation involves several steps: first, the log-likelihood is computed for a range of $H_0$ values using the theoretical dispersion measure model $\text{DM}_{\text{model}}(z, H_0)$, which incorporates contributions from the intergalactic medium, host galaxies, and the Milky Way. Uncertainties are propagated through a quadratic sum to account for variability in each component:
\begin{align}
\sigma_{\text{i}} = \sqrt{\sigma_{\text{MW}}^2(z_i) + \sigma_{\text{host}}^2(z_i) + \sigma_{\text{LSS}}^2(z_i)}.
\end{align}

The value of $H_0$ that maximizes the log-likelihood corresponds to the most probable estimate of the Hubble constant given the data, and it is identified by locating the peak of the likelihood function. The uncertainty in $H_0$ is estimated by calculating the second derivative of the log-likelihood at its maximum, with the error given by:
\begin{equation}
\sigma_{H_0} = \sqrt{\frac{1}{-\ln \mathcal{L}''(H_{0;\text{best}})}},
\end{equation}
\noindent where $\ln \mathcal{L}''(H_{0;\text{best}})$ represents the second derivative of the log-likelihood evaluated at the maximum.\newline

To incorporate theoretical constraints, a rectangular prior is assumed in a range of $[40, 100]$ for $H_0$, assigning a uniform probability density within this interval. The posterior distribution is constructed by multiplying the likelihood function by this prior. The peak of the posterior represents the combined (best) estimate of $H_0$. Using the MLE method, the best fit value for $H_0$ is:
\begin{align}\label{H0Likelihood}
H_0 &= 65.13 \pm 2.52\,\text{km/s/Mpc}.
\end{align}
The value reported in \eqref{H0Likelihood} has a statistical precision of 3.9\%, significantly better than that obtained in the previous section using the arithmetic mean. Thus, our implementation of the MLE exhibits superior performance compared with the methods discussed in previous sections. Finally, the estimated value of $H_0$ is compatible with the early universe prediction \cite{planck2018} and disfavors the predicted value by \cite{riess2022}.
\begin{figure}
\begin{center}
\includegraphics[width=9cm]{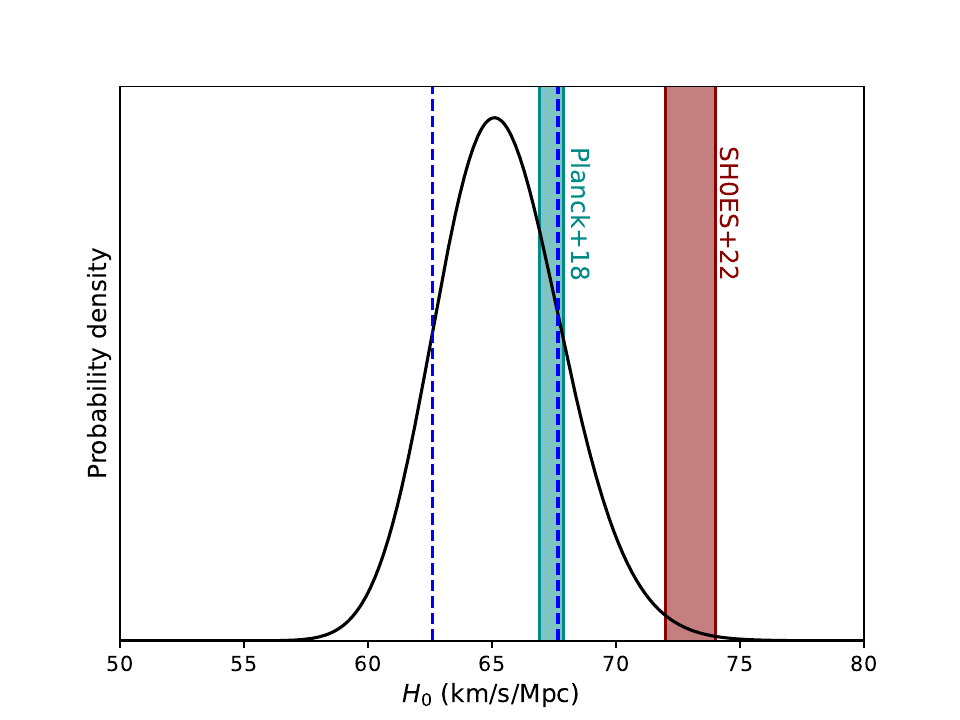}
\end{center}
\caption{Posterior probability density function for $H_0$ with 97 localized FRBs. Our best-fit value with the MLE method is $H_0 = 65.13 \pm 2.52\,\text{km/s/Mpc}$. Within the error margin (blue dotted lines), this prediction is compatible with other indirect methods, such as the one presented in \citet{planck2018}.}
\end{figure}

\section{Future prospectives with FRBs mock data}\label{mocks} 
Currently, the available set of confirmed FRBs is limited; however, it is possible to simulate data to gain insight into the potential of these methods.\newline

Following the procedure proposed in \cite{YuWang2017, Liu_2023, fortunato2024}, we adopt a flat $\Lambda$CDM model as the fiducial model, with the following cosmological parameters: $H_0 = 70\, \text{km/s/Mpc}$, $\Omega_b = 0.049$, $\Omega_m = 0.3$, and $\Omega_\Lambda = 1 - \Omega_m$.  Furthermore, we assume that the redshift distribution is described by $f(z) \approx z^2 \exp(-\alpha z)$ , with  $\alpha = 7$  \cite{hagstotz2022}, a parameter that determines the depth of the simulated sample, acting as a filter that places most of the data points within the range $z = 0.15 - 0.5$ consistent with confirmed data to-date.\newline
The $\text{DM}_\text{IGM}$ term is proposed through a normal distribution 
$N(\langle \text{DM}_{\text{IGM}}^{\text{fid}} \rangle, \sigma_{\text{DM}_{\text{IGM}}^{\text{fid}}}) $, where $ \sigma_{\text{DM}_{\text{IGM}}^{\text{fid}}} = \sigma_{\Delta \text{IGM}}(z)$. The term \( \langle \text{DM}_{\text{IGM}}^{\text{fid}} \rangle \) is calculated in eq.~\eqref{IGM}. The large dispersion of the values of $\text{DM}_\text{IGM}$ around its mean value $\langle\text{DM}_\text{IGM}\rangle$ is due to the inhomogeneity of the distribution of baryons in the intergalactic medium is modeled by with a power-law function presented in \cite{McQuinn_2014, Qiang_2020, fortunato2024}:
\begin{equation}
\sigma_{\Delta \text{IGM}}(z) = 173.8 \, z^{0.4} \, \text{pc} \, \text{cm}^{-3}.
\end{equation}
The contribution of the host galaxy is described by a log-normal distribution, such that:
\begin{equation}
P(\text{DM}_{\text{host}}) = \frac{1}{\text{DM}_{\text{host}} \, \sigma_{\text{host}} \sqrt{2\pi}} 
\exp \left( -\frac{\left( \ln(\text{DM}_{\text{host}} / \mu) \right)^2}{2 \sigma_{\text{host}}^2} \right).
\end{equation}
Here, $\mu$ is the geometric mean, and we consider it within the range $[40, 80] \, \text{pc\,cm}^{-3}$. The term $\sigma_{\text{host}}$ is
 within the interval $[0.4, 1.0]$.\newline
 
\begin{figure}
\begin{center}
\includegraphics[width=9cm]{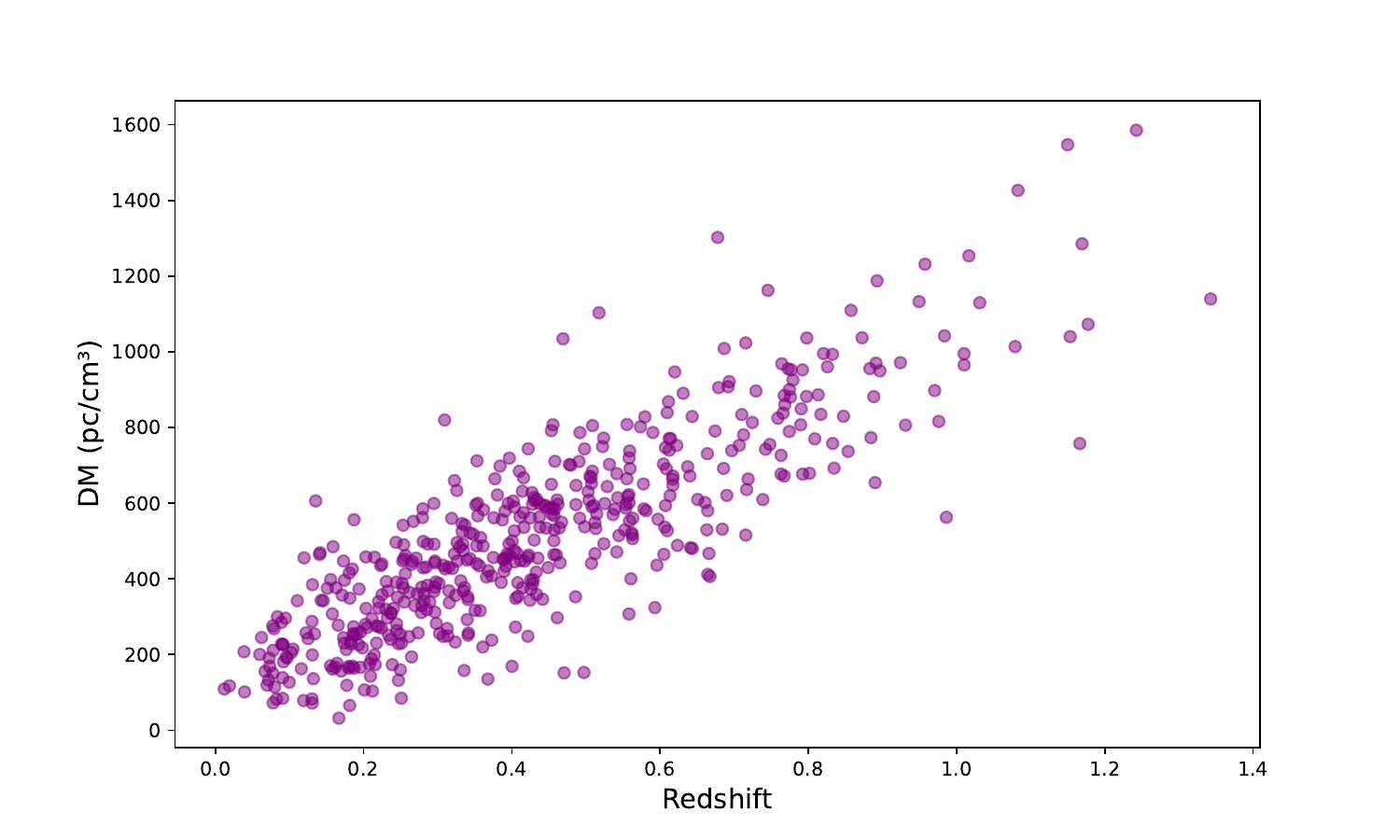}
\end{center}
\caption{\tiny{Reconstruction of the dispersion measure and redshift for one realization out of the 100 mock datasets. }}\label{datasample}
\end{figure}

 We generate 100 independent realizations from random seeds between 0 and 199. Each set contains 500 simulated FRBs. We display one example of these synthetic samples in Figure~\ref{datasample}. For each realization, we calculate a value of $H_0$ applying the methods described in previous sections, such that we recover a set of 100 different values for $H_0$. Finally, we calculate the arithmetic mean and standard deviation for all these $H_0$ values, which yields the results shown in Table~\ref{resultMock}.\newline

 The arithmetic mean is highly sensitive to the tails of the data distribution. In contrast, an analysis using the median proves to be more robust with an improvement in a factor of $\sim 2$ in the statistical precision. The median value is only computed for our simulated dataset since it has proven this metric is more robust with large datasets. The median leads to a value for the Hubble parameter today of $66.10\pm 1.89 \, \text{km/s/Mpc}$, which is highly consistent with the prediction of \cite{planck2018} but with lower statistical precision (2.9\%) compared to arithmetic mean. The error reported for the median is the median absolute deviation (MAD).

\begin{table}
\centering
\begin{tabular}{|l|l|c|c|}
\hline
\multicolumn{2}{|l|}{\textbf{Method}}                                                                                          & $H_0$   & Statistical \\ 
\multicolumn{2}{|l|}{}                                                                                          &  $(\text{km/s/Mpc})$ & precision ($\%$) \\ \hline
\multicolumn{2}{|l|}{Arithmetic mean}                                                                                           & $66.21\pm 3.46$ &    5.2         \\ \hline
\multicolumn{2}{|l|}{Median}                                                                                                   & $66.10 \pm 1.89$  &    2.9        \\ \hline
\multicolumn{2}{|l|}{MLE}                                                                                               & $67.30 \pm 0.91$  &      1.4      \\ \hline
\multirow{2}{*}{$H(z)$} & Linear model      & $54.34\pm 1.57$ & 2.9 \\ \cline{2-4} 
                        & Power-law model & $91.84\pm 1.82$ & 2.0 \\ \hline
\end{tabular}
\caption{\tiny{Value of $H_0$ in the different methods using 100 mock catalogs with 500 synthetic data each.}}\label{resultMock}
\end{table}

For all methods applied in synthetic data, the values of $H_0$ increase compared with those found with our catalog of 98 observed FRBs, suggesting that with the current detected set of transients, the value of the Hubble constant might be underestimated, as indicated by the results in this section. The statistical precision will continue to improve as more confirmed data become available. On the other hand,  our prediction using the arithmetic mean as a metric remains considerably high compared to the other methods, indicating that it would require more data to reliably estimate $H_0$. The best performance is achieved through the implementation of the MLE method, with values from $65.13 \pm 2.52 \, \text{km/s/Mpc}$ (statistical precision $\sim$3.9\%) calculated with the 97 FRBs in Table~\ref{data} and $67.30 \pm 0.91 \, \text{km/s/Mpc}$ (statistical precision 1.4\%) for our synthetic dataset. In summary, the statistical precision improves by a factor of 2.8 when new data is used as input in our calculations. The statistical precision in the mock-case is comparable to the precision reported by the SH0ES collaboration. On the other hand, methods based on the reconstruction of $H(z)$ exhibit the largest deviations from the values reported in \cite{planck2018} and \cite{riess2022}. Notably, our prediction with the median value has a statistical precision similar to the methods based on $H(z)$ but it is compatible with the value of $H_0$ reported by \cite{planck2018}.

\section{Conclusions}\label{sec:concl}
The Hubble constant ($H_0$) is a very important cosmological parameter that measures the rate of expansion of the universe at current times. The so-called ``Hubble tension" is the discrepancy with a gap of more than $5\sigma$ between the predictions of $H_0$ values by early \cite{planck2018} and late \cite{riess2022} universe observations. Fast radio bursts (FRBs) offer new pathways to quantify this cosmological parameter. In particular, it can be used to address the Hubble tension based on the sensitivity of the dispersion measure, in particular, in the IGM term. We construct a data set with 98 localized FRBs from literature and explore three different methods to estimate $H_0$ value with 97 of them: i) arithmetic mean, ii) reconstruction of $H(z)$ and iii) the Maximum Likelihood Estimate. \newline 
In the first method, based on the sensitivity of the intergalactic medium's dispersion measure ($\text{DM}_{\text{IGM}}$) to the Hubble constant, $H_0$  is calculated using the 97 FRB data points in the catalog and its errors (Table~\ref{data}). Using these data, we calculate a representative value as the arithmetic mean, with its associated error estimated through error propagation, yielding $H_0 = 57.67\pm 11.99 \, \text{km/s/Mpc}$, with a statistical precision of 20.7\%. This result shows agreement with results from \cite{planck2018}.\newline

In the second method, the cosmic expansion history $H(z)$ is reconstructed by taking the derivative of the average value  $\text{DM}_{\text{IGM}}$ with respect to $z$. Two DM-$z$ relations are considered for calculating the derivative: a linear and a power-law function. These relationships are fitted using the 97 confirmed data points. Once $H(z)$ is reconstructed, we set $z=0$ to recover $H_0$. The linear function leads to a value of $H_0$ of $51.27 ^{+3.80}_{-3.31}  \, \text{km/s/Mpc}$, closer to reports by \cite{planck2018}, with a statistical precision of 7.4\%. On the other hand, unlike previous results, the power-law model predicts a higher value than \cite{riess2022},  $77.09^{+8.89}_{-7.64}  \, \text{km/s/Mpc}$. However, it exhibits lower statistical precision than the linear model, at 11.5\% level. \newline
Finally, we use the MLE method to determine the best value of $H_0$ that fits the proposed model to the 97 localized FRBs. This method achieves the best statistical precision at 3.9\%, predicting a value for the Hubble parameter today of $H_0 = 65.13 \pm 2.52  \, \text{km/s/Mpc}$. The latter result is compatible with reports that use data from the early universe\cite{planck2018}.\newline
Overall, our results with observed data have lower statistical precision compared to \cite{planck2018,riess2022}. However, the number of FRBs with confirmed host galaxies is expected to increase significantly shortly. Therefore, it seems promising to evaluate these methods with larger datasets. To this end, we generate 100 mock catalogs with 500 data points each. We assume a particular redshift distribution discussed in section \ref{mocks}, a normal distribution for DM$_{\text{IGM}}$ and we take a log-normal distribution for the host galaxy dispersion measure and repeat the calculations with the three methods discussed with 500 data points to recover a set of 100 different values for $H_0$. Finally, we calculate the arithmetic mean for each method and obtain the following results for $H_0$:  $66.21\pm 3.46\, \text{km/s/Mpc}$, $66.10\pm 3.46\, \text{km/s/Mpc}$, $67.30\pm1.89\, \text{km/s/Mpc}$, $54.34\pm1.57\, \text{km/s/Mpc}$ and $91.84\pm1.82\, \text{km/s/Mpc}$ for arithmetic mean, median,  MLE method and the $H(z)$ reconstruction method with a linear and a power-law function, respectively. In all the methods applied to mock data, the statistical precision improves significantly, but the MLE method stands out with a value of 1.4\%, which is in the order of the reports by \cite{riess2022}.\newline

Besides the values reported for $H_0$ with the different methods explored in this work, we highlight that the statistical precision of the predicted value of $H_0$ increases by a factor of 2 if the set of confirmed (and localized) FRBs grows at least 5 times the current sample. More importantly, our results are consistent with other indirect methods to infer the Hubble factor today since the main observable considered here, the dispersion measure (DM), is effectively a distance from the observer and the FRB. Hence, it is unsurprising that most of our predictions lie in the range of values predicted with other indirect methods, such as CMB or BAO. Although our sample of FRBs remains in a low redshift regime ($z_{\text{high}} \sim$ 1.3), our results are more compatible with predictions of $H_0$ derived with early universe probes than forecasts of this cosmological parameter made with observations in the local universe.

\section*{Conflict of Interest Statement}
The authors declare that the research was conducted in the absence of any commercial or financial relationships that could be construed as a potential conflict of interest.

\section*{Funding}

This work was supported by Fundación Universitaria Los Libertadores programme ``Décimo Segunda (XII) Convocatoria Interna Anual de Proyectos de Investigación, Creación Artística y Cultural", project ``Estimación del espacio de parámetros para modelos difusivos cosmológicos a través de métodos bayesianos y de machine learning." [Grant number: ING-14-24].

\section*{Acknowledgments}
The authors thank Fundaci\'on Universitaria Los Libertadores and Universidad ECCI for granting us the resources to develop this project. This material is based upon work supported by the Google Cloud Research Credits program with the award GCP19980904. This research made use of \textsc{matplotlib} \cite{hunter2007}, \textsc{SciPy} \cite{virtanen2020} and \textsc{NumPy} \cite{harris2020}.

\section*{Data Availability}

The inclusion of a Data Availability Statement is a requirement for articles published in MNRAS. Data Availability Statements provide a standardised format for readers to understand the availability of data underlying the research results described in the article. The statement may refer to original data generated in the course of the study or to third-party data analysed in the article. The statement should describe and provide means of access, where possible, by linking to the data or providing the required accession numbers for the relevant databases or DOIs.



\bibliographystyle{mnras}
\bibliography{mnras} 



\onecolumn
\appendix

\section{Our working catalog}\label{appen}

The data presented in Table~\ref{data} are a collection of 98 Fast Radio Bursts (FRBs) detected at various coordinates across the sky. Each entry provides detailed information about the event's position and signal characteristics. Right Ascension (RA) and Declination (DEC) are expressed in degrees, indicating the precise sky location of each event. The observed dispersion measure (DM$_{\text{obs}}$), measured $\text{pc\,cm}^{-3}$, quantifies the amount of ionized material the signal has traveled through and is accompanied by its associated error ($\Delta$DM) to indicate the precision of the measurement. The table also specifies whether the FRB is a repeater (Rep), with ``Y" for repeaters and ``N" for single occurrences. Additionally, the redshift ($z_{\text{host}}$) of the host galaxy is included, providing an estimate of the cosmological distance to the event, along with its error margin ($\Delta z_{\text{host}}$). Lastly, the bibliographic reference for each detection is listed. FRB 221027A is excluded from our calculations due to ambiguity in its host galaxy localization\newline

\begin{longtable}{ccccccccl}
    \caption{98 localized FRBs and their properties} \label{data} \\
    \toprule
        FRB & RA  & DEC & DM$_{\text{obs}}$ & $\Delta\text{DM}$ & Rep & $z_{\text{host}}$ & $\Delta z_{\text{host}}$ & Reference \\
         & (deg) & (deg) & (pc\,cm$^{-3}$) & (pc\,cm$^{-3}$) &  &  &  &  \\
        \midrule
         121102A  & 82.9946  & 33.1479  & 557    & 2    & Y  & 0.1927  & -     &  \citeyear{Petroff2016}  \\
         171020A  & 333.75   & -19.6667 & 114.1  & 0.2  & N  & 0.008672 & -     & \citeyear{Leewaddell2023} \\
         180301A  & 93.2292  & 4.6711   & 536    & 5    & Y  & 0.3305  & -     & \citeyear{Bhandari2022} \\
         180814A  & 65.68    & 73.66    & 189.4  & 3.23 & Y  & 0.068   & -     &  \citeyear{Michilli_2023}  \\
         180916B  & 29.5031  & 65.7168  & 348.8  & 1.62 & Y  & 0.0337  & -     & \citeyear{Heintz2020} \\
         180924B  & 326.1052 & -40.9    & 362.16 & 0.06 & N  & 0.3214  & -     & \citeyear{Heintz2020, Gordon_Fong_2023} \\
         181030A  & 163.2    & 73.74    & 103.5  & 1.62 & Y  & 0.0039  & -     & \citeyear{Bhardwaj2021} \\
         181112A  & 327.3485 & -52.9709 & 589    & 0.03 & N  & 0.4755  & -     & \citeyear{Heintz2020} \\
         181220A  & 348.6982 & 48.3421  & 208.66 & 1.62 & N  & 0.02746 & -     & \citeyear{Bhardwaj2023}  \\
         181223C  & 180.9207 & 27.5476  & 111.61 & 1.62 & N  & 0.03024 & -     & \citeyear{Bhardwaj2023}  \\
         190102C  & 322.4157 & -79.4757 & 364.545 & 0.3  & N  & 0.2913  & -     & \citeyear{Heintz2020} \\
         190110C  & 249.3185 & 41.4434  & 221.6  & 1.62 & Y  & 0.12244 & -     & \citeyear{Ibik_2024} \\
         190303A  & 207.9958 & 48.1211  & 223.2  & 1.62 & Y  & 0.064   & -     &  \citeyear{Michilli_2023}  \\
         190418A  & 65.8123  & 16.0738  & 182.78 & 1.62 & N  & 0.07132 & -     & \citeyear{Bhardwaj2023}  \\
         190425A  & 255.6625 & 21.5767  & 127.78 & 1.62 & N  & 0.03122 & -     & \citeyear{Bhardwaj2023}  \\
         190520B  & 240.5167 & -11.2883 & 1204.7 & 4    & Y  & 0.241   & 0.001 & \citeyear{Niu2022, Gordon_Fong_2023} \\
         190523A  & 207.065  & 72.4697  & 760.8  & 0.6  & N  & 0.66    & 2     &  \citeyear{Ravi2019}   \\
         190608B  & 334.0199 & -7.8982  & 340.05 & 0.5  & N  & 0.1178  & -     & \citeyear{Heintz2020, Gordon_Fong_2023} \\
         190611B  & 320.7455 & -79.3976 & 332.63 & 0.2  & N  & 0.3778  & -     & \citeyear{Heintz2020} \\
         190614D  & 65.07552 & 73.70674 & 959.2  & 0.5  & N  & 0.6     & -     & \citeyear{Law2020ApJ} \\
         190711A  & 329.4195  & -80.358   & 592.6   & 0.4  & Y  & 0.5217  & -     & \citeyear{Heintz2020} \\
         190714A  & 183.9797  & -13.021   & 504.13  & 2    & N  & 0.2365  & -     & \citeyear{Heintz2020} \\
         191001A  & 323       & -54.6667  & 507.9   & 0.04 & N  & 0.234   & -     & \citeyear{Petroff2016}  \\
         191106C  & 199.5801  & 42.9997   & 332.2   & 0    & Y  & 0.10775 & -     & \citeyear{Ibik_2024} \\
         191228A  & 344.4292  & -29.5942  & 297.5   & 0.05 & N  & 0.2432  & -     & \citeyear{Bhandari2022} \\
         200120E  & 146.25    & 68.77     & 87.82   & 1.62 & Y  & 0.0008  & -     & \citeyear{Nimmo2023} \\
         200223B  & 82.695    & 288.313   & 201.8   & -    & Y  & 0.06024 & -     & \citeyear{Ibik_2024} \\
         200430A  & 229.7064  & 12.3769   & 380.1   & 0.4  & N  & 0.1608  & -     & \citeyear{Heintz2020} \\
         200906A  & 53.4958   & -14.0831  & 577.8   & 0.2  & N  & 0.3688  & -     & \citeyear{Bhandari2022} \\
         201123A  & 263.669   & -50.7672  & 433.55  & -    & Y  & 0.0507  & -     & \citeyear{Rajwade2022}  \\
         201124A  & 76.99     & 26.19     & 413.52  & 3.23  & Y  & 0.0979  & -      & \citeyear{Wangfy2022}  \\
         210117A  & 339.9792  & -16.1517  & 728.95  & 0.36  & N  & 0.214   & 0.001  & \citeyear{Bhandari2023, Gordon_Fong_2023}  \\
         210320C  & 204.32    & -15.4104  & 384.8   & 0.3   & N  & 0.2797  & -      & \citeyear{James2022MNRAS, Gordon_Fong_2023}  \\
         210405I  & 255.3396  & -49.5452  & 565.17  & -     & N  & 0.066   & -      & \citeyear{Driessen2023}    \\
         210410D  & 326.0863  & -79.3182  & 578.78  & -     & N  & 0.1415  & -      & \citeyear{Gordon_Fong_2023, Caleb2023}    \\
         210603A  & 10.274    & 21.226    & 500.147 & 0.004 & N  & 0.1772  & 0.0001 & \citeyear{NatureCassanelli2024}  \\
         210807D  & 299.2042  & -0.8143   & 251.9   & 0.2   & N  & 0.12927 & -      & \citeyear{James2022MNRAS, Gordon_Fong_2023}   \\
         211127I  & 199.7896  & -18.8246  & 234.83  & 0.08  & N  & 0.0469  & -      & \citeyear{James2022MNRAS, Gordon_Fong_2023}   \\
         211203C  & 204.47    & -31.3678  & 636.2   & 0.4   & N  & 0.34386 & -      & \citeyear{ Gordon_Fong_2023, Shannon2024Baptista2023}  \\
         211212A  & 157.6696  & 1.6769    & 206     & 5     & N  & 0.0715  & -      & \citeyear{James2022MNRAS, Gordon_Fong_2023}   \\
         220105A  & 208.9642  & 22.4888   & 583     & 1    & N  & 0.2785  & -      & \citeyear{ Gordon_Fong_2023,Shannon2024Baptista2023}  \\
         220204A  & 278.3321  & 71.6157   & 612.2   & -    & N  & 0.4     & -      & \citeyear{Sherman_2024, sharma2024preferentialoccurrencefastradio}   \\
         220207C  & 310.1995  & 72.8823   & 263     & -    & N  & 0.043   & -      & \citeyear{Law2023}  \\         
         220208A  & 319.3483  & 71.54     & 437     & -    & N  & 0.351   & -      & \citeyear{sharma2024preferentialoccurrencefastradio}   \\

    \endfirsthead
    
    \multicolumn{9}{c}{{\tablename\ \thetable{} -- continued from previous page}} \\
    \toprule
        FRB & RA  & DEC & DM$_{\text{obs}}$ & $\Delta\text{DM}$ & Rep & $z_{\text{host}}$ & $\Delta z_{\text{host}}$ & Reference \\
     & (deg) & (deg) & (pc\,cm$^{-3}$) & (pc\,cm$^{-3}$) &  &  &  &  \\
        \midrule
        \endhead
         220307B  & 350.8745  & 72.1924   & 499.328 & -    & N  & 0.248   & -      & \citeyear{Law2023}  \\
         220310F  & 134.7205  & 73.4908   & 462.657 & -    & N  & 0.478   & -      & \citeyear{Law2023}  \\
         220319D  & 32.1779   & 71.035    & 110.95  & 0.01 & N  & 0.011   & -      & \citeyear{Law2023}  \\
         220330D  & 165.7256  & 71.7535   & 468.1   & -    & N  & 0.3714  & -      & \citeyear{sharma2024preferentialoccurrencefastradio}   \\
         220418A  & 219.1056  & 70.0959   & 624.124 & -    & N  & 0.622   & -      & \citeyear{Law2023}  \\
         220501C  & 352.3792  & -32.4907  & 449.5   & 0.2  & N  & 0.381   & -      & \citeyear{Driessen2023, shannon2024, sharma2024preferentialoccurrencefastradio, Gao_2024kkx} \\
         220506D  & 318.0448  & 72.8273   & 396.651  & -     & N  & 0.3     & -      & \citeyear{Law2023, sharma2024preferentialoccurrencefastradio}  \\
         220509G  & 282.67    & 70.2438   & 269.53   & 10    & N  & 0.0894  & -      & \citeyear{Connor2023}  \\
         220529A  & 19.10     & 20.63     & 246      & -     & Y  & 0.1839  & -      & \citeyear{Gao_2024kkx}   \\
         220610A  & 351       & -33.5167  & 1458.1   & 0.2   & N  & 1.016   & 0.002  & \citeyear{Gordon2023}     \\
         220717A  & 293.3042  & -19.2877  & 637.34   & -     & N  & 0.36295 & -      & \citeyear{Rajwade2024} \\
         220725A  & 336.75    & 34.8833   & 290.4    & 0.3   & N  & 0.1926  & -      & \citeyear{shannon2024}  \\
         220726A  & 75.1058   & 71.6018   & 686.55   & -     & N  & 0.361   & -      & \citeyear{ sharma2024preferentialoccurrencefastradio}  \\
         220825A  & 311.9815  & 72.5850   & 649.893  & -     & N  & 0.241   & -      & \citeyear{connor2024gasrichcosmicweb}   \\
         220831A  & 333.0854  & 71.5376   & 1146.25  & -     & N  & 0.262   & -  &  \citeyear{connor2024gasrichcosmicweb}   \\
         220912A  & 347.2704  & 48.7071   & 219.46   & 0.042 & Y  & 0.0771  & -      & \citeyear{Zhang2023}    \\
         220914A  & 282.0568  & 73.3369   & 631.29   & 10    & N  & 0.1139  & -      & \citeyear{Connor2023} \\
         220918A  & 17.7413   & -70.785  & 657     & 0.4   & N  & 0.491  & -      & \citeyear{Shannon2024Baptista2023}  \\
         220920A  & 240.2571  & 70.9188   & 314.977 & -    & N  & 0.158  & -      & \citeyear{Law2023}  \\
         221012A  & 280.7987  & 70.5242   & 440.358 & -     & N  & 0.285  & -      & \citeyear{Law2023}  \\
         221027A  & 129.6104  & 71.7315   & 452.5   & -     & N  & 0.229/0.5422  & -      &  \citeyear{connor2024gasrichcosmicweb} \\
         221029A  & 143.8351  & 71.7529   & 1391.05 & -     & N  & 0.975  & -      & \citeyear{ sharma2024preferentialoccurrencefastradio} \\
         221101B  & 341.4589  & 71.5295   & 490.7   & -     & N  & 0.2395 & -      & \citeyear{ sharma2024preferentialoccurrencefastradio}\\
         221106A  & 56.7048   & -25.5698  & 343.8   & 0.8   &  N  & 0.2044 & -      & \citeyear{Sherman_2024,shannon2024} \\
         221113A  & 72.8406   & 71.6131   & 411.4   & -     & N  & 0.2505 & -      & \citeyear{ sharma2024preferentialoccurrencefastradio}  \\
         221116A  & 17.6617   & 71.5288   & 640.6   & -     & N  & 0.2764 & -      &   \citeyear{ sharma2024preferentialoccurrencefastradio}  \\
         221219A  & 255.7773  & 71.6817   & 706.7   & -   & N  & 0.554  & -      &  \citeyear{ sharma2024preferentialoccurrencefastradio} \\
         230124A  & 233.0768  & 71.7273   & 590.6   & -   & N  & 0.094  & -      & \citeyear{ sharma2024preferentialoccurrencefastradio}  \\
         230216A  & 155.9717  & 1.4678    & 828     & -   & N  & 0.531  & -      & \citeyear{shannon2024, connor2024gasrichcosmicweb}   \\
         230307A  & 177.78    & 71.41     & 608.9   & -   & N  & 0.2710 & -      & \citeyear{ sharma2024preferentialoccurrencefastradio}\\
         230501A  & 338.5535  & 71.5292   & 532.5   & -   & N  & 0.301  & -      & \citeyear{Sherman_2024, connor2024gasrichcosmicweb}   \\
         230521B  & 349.6785  & 71.5220   & 1342.9  & -   & N  & 1.354  & -      & \citeyear{shannon2024, connor2024gasrichcosmicweb}  \\
         230526A  & 22.3646   & -52.7688  & 316.4   & 0.2 & N  & 0.157  & -      & \citeyear{shannon2024}   \\
         230626A  & 240.7125  & 71.7142   & 451.2   & -   & N  & 0.327  & -      & \citeyear{sharma2024preferentialoccurrencefastradio} \\
         230628A  & 161.8999  & 71.7745   & 345.15  & -   & N  & 0.1265 & -      & \citeyear{ sharma2024preferentialoccurrencefastradio}  \\
         230708A  & 303.2371  & -55.3807  & 411.5   & 0.06 & N  & 0.105  & -      & \citeyear{shannon2024}   \\
         230712A  & 170.7112  & 71.7794   & 586.96  & -   & N  & 0.4525 & -      & \citeyear{ sharma2024preferentialoccurrencefastradio}  \\
         230718A  & 127.6129  & -41.0036  & 477     & 0.5 & N  & 0.035  & -      & \citeyear{shannon2024}   \\
         230814A  & 335.9746  & 73.0259   & 696.4   & 0.05 & N  & 0.5535 & -      & \citeyear{connor2024gasrichcosmicweb} \\
         230902A  & 52.3671   & -47.5626  & 440.1   & 0.1  & N  & 0.3619 & -      & \citeyear{shannon2024}  \\
         231120A  & 143.6169  & 71.7574   & 438.9   & -   & N  & 0.07   & -      &  \citeyear{ sharma2024preferentialoccurrencefastradio} \\
         231123B  & 240.5665  & 71.7156   & 396.7   & -   & N  & 0.2625 & -      & \citeyear{ sharma2024preferentialoccurrencefastradio} \\
         231220A  & 122.2054  & 71.7217   & 491.2   & -   & N  & 0.3355 & -      & \citeyear{connor2024gasrichcosmicweb} \\
         231226A  & 155.2817  & 6.1294    & 329.9   & 0.1 & N  & 0.1569 & -      & \citeyear{shannon2024}   \\
         240114A  & 322.0703  & 4.4841    & 527.7   & -   & Y  & 0.13   & -      &  \citeyear{Tian2024}\\
         240119A  & 218.1169  & 71.7554   & 483.1   & -   & N  & 0.37   & -      &   \citeyear{connor2024gasrichcosmicweb}\\
         240123A  & 66.134   & 71.5965   & 1462    & -   & N  & 0.968   & -      &   \citeyear{connor2024gasrichcosmicweb} \\
         240124A  & 321.9162  & 4.3501    & 526.9   & -   & Y  & 0.269   & 0.139  & \citeyear{Duncan2022}  \\
         240201A  & 149.9056  & 14.088   & 374.5   & 0.3 & N  & 0.042729  & -      & \citeyear{shannon2024} \\
         240210A  & 8.7796    & -28.2708  & 283.73  & 0.05 & N  & 0.0237686 & -      & \citeyear{shannon2024} \\
         240213A  & 158.7613  & 9    & 357.4   & -   & N  & 0.1185  & -      & \citeyear{connor2024gasrichcosmicweb} \\
         240215A  & 268.4333  & 71.6540   & 549.5   & -   & N  & 0.21    & -     &\citeyear{connor2024gasrichcosmicweb}  \\
         240229A  & 173.7346  & 71.7838   & 491.15  & -   & N  & 0.287   & -      & \citeyear{connor2024gasrichcosmicweb} \\
         240310A  & 17.6219   & -44.4394  & 601.8   & 0.2 & N  & 0.127   & -      & \citeyear{shannon2024} \\
\end{longtable}


\bsp	
\label{lastpage}
\end{document}